\documentclass[sigconf, nonacm]{acmart}

\usepackage{multirow}
\usepackage{xspace}
\usepackage{xcolor}

\usepackage{pgfplots}
\usepackage{pgfplotstable}
\pgfplotsset{compat=1.18}
\usepackage{booktabs}
\usepackage{subcaption}
\usepackage{tabularx}
\usepackage{siunitx}  
\usepackage{adjustbox}

\newcommand\vldbyear{2025}
\newcommand\vldbworkshop{14th International Workshop on Quality in Databases (QDB'25)}
\newcommand\vldbauthors{\authors}
\newcommand\vldbtitle{\shorttitle} 
\newcommand\vldbavailabilityurl{URL_TO_YOUR_ARTIFACTS}
\newcommand\vldbpagestyle{plain} 

\begin{document}
\title{PBE Meets LLM: When Few Examples Aren’t Few-Shot Enough}

\author{Shuning Zhang}
\affiliation{%
  \institution{University of Illinois Urbana-Champaign}
}
\email{sz31@illinois.edu}
\thanks{Work done while Shuning Zhang was at the University of Illinois, Urbana-Champaign. The author is now with Meta.}

\author{Yongjoo Park}
\affiliation{%
  \institution{University of Illinois Urbana-Champaign}
}
\email{yongjoo@illinois.edu}

\begin{abstract}

Large language models (LLMs) can generate code from natural language descriptions. Their performance is typically evaluated using programming benchmarks that simulate real-world tasks. These benchmarks provide specifications in the form of docstrings, function signatures, or bug reports. The model then generates a program, which is tested against predefined test cases. In contrast, Programming by Example (PBE) uses input-output examples as the specification. Traditional PBE systems rely on search-based methods over restricted transformation spaces. They are usually designed for narrow domains and fixed input formats. It remains unclear how well LLMs perform on PBE tasks.

In this work, we evaluate LLMs on PBE tasks involving tabular data transformations. We prompt models to generate functions that convert an input table to an output table. We test the generated functions on unseen inputs to measure accuracy. Our study includes multiple LLMs and evaluates different prompting strategies, such as one-shot vs.~multi-try.
We also compare performance with and without PBE-specific knowledge. 
Finally, we propose a hybrid method that calls a traditional PBE solver first, and then falls back to LLMs if necessary. 
Our results show that LLMs support more diverse input formats and achieve higher accuracy than conventional methods. However, they struggle with tasks that contain ambiguity. The hybrid approach improves overall success by combining the strengths of both approaches.

\end{abstract}

\maketitle

\pagestyle{\vldbpagestyle}
\begingroup\small\noindent\raggedright\textbf{VLDB Workshop Reference Format:}\\
\vldbauthors. \vldbtitle. VLDB \vldbyear\ Workshop: \vldbworkshop.\\ 
\endgroup
\begingroup
\renewcommand\thefootnote{}\footnote{\noindent
This work is licensed under the Creative Commons BY-NC-ND 4.0 International License. Visit \url{https://creativecommons.org/licenses/by-nc-nd/4.0/} to view a copy of this license. For any use beyond those covered by this license, obtain permission by emailing \href{mailto:info@vldb.org}{info@vldb.org}. Copyright is held by the owner/author(s). Publication rights licensed to the VLDB Endowment. \\
\raggedright Proceedings of the VLDB Endowment. 
ISSN 2150-8097. \\
}\addtocounter{footnote}{-1}\endgroup

\ifdefempty{\vldbavailabilityurl}{}{
\vspace{.3cm}
\begingroup\small\noindent\raggedright\textbf{VLDB Workshop Artifact Availability:}\\
The source code, data, and/or other artifacts have been made available at \url{https://github.com/illinoisdata/PBE-Meets-LLM.git}.
\endgroup
}

\section{Introduction}

Modern large language models (LLMs) \cite{gpt4, llama3} are capable of generating high-quality programs from natural language descriptions. To assess their performance, researchers commonly use a variety of coding benchmarks that simulate real-world programming tasks. These benchmarks provide models with different forms of program specifications, such as docstrings and function signatures~\cite{humaneval}, function-level descriptions~\cite{apps, mbpp, bigcodebench}, class-level descriptions~\cite{classeval}, or even bug reports~\cite{jimenez2023swe, issue2test}. The model then generates a corresponding program, which is subsequently evaluated based on its ability to pass predefined test cases.

However, LLMs are less evaluated for a different mode of programming, i.e., Programming by Example (PBE) \cite{pbe_1st,pbe_2nd}.
Using tabular data as input-output examples as program specifications, PBE generates the best-suited program, which can be replicated on other datasets.
PBE has been developed for number/string manipulation \cite{StringProcessing, SemanticString, NumberTransformations, Wang2017SynthesizingSQL, string} and tabular data transformation (e.g., FlashFill \cite{flashfill} and FlashExtract \cite{flashextract}).
These conventional PBE systems rely on techniques very different from how LLMs generate code.
PBE methods perform a search (e.g., A* search \cite{foofah}) within a more restrictive space of data transformation patterns, such as moving rows/columns, combining cells, and so on \cite{potterswheel}.
It is not straightforward whether LLMs, trained on massive datasets, would generate code that resembles the one produced by conventional PBE methods, or whether they would generate a completely different program.

In this work,
we evaluate LLMs on a rerely explored area PBE tasks---tabular data 
    transformations---to assess their generality and accuracy.
That is,
we construct prompts that ask the model to generate a function
that transforms an example input table into a corresponding output table.
The input table is provided in JSON format, capturing its structure.
Then, we test the generated function
on unseen test input tables
to determine whether the desired output tables are produced.
Our evaluation uses the latest proprietary LLMs (e.g., OpenAI GPT-4o)
to compare their performance and identify limitations
across a wide range of PBE tasks curated from prior work.
Moreover, we examine
whether these LLMs can achieve higher performance
with different techniques,
such as (1) one-shot vs multi-tries, 
(2) no external knowledge vs additional knowledge, 
and (3) a hybrid framework combining LLMs with traditional PBE models.

Our study shows that
modern LLMs can achieve high accuracy on PBE tasks,
covering more diverse input formats
compared to conventional PBE methods,
which are typically designed for a specific input format.
For example,
a conventional PBE method, FooFah \cite{foofah},
demonstrates good performance on its curated datasets,
but fails more frequently on datasets prepared by another paper (e.g., Prose \cite{prose}).
The reverse is also true.
In contrast, LLMs tend to show high accuracy
across diverse benchmark sets.
However,
our study also reveals limitations of LLMs.
They often produce incorrect programs when the task contains inherent ambiguity,
even if the correct transformation might be understandable to human engineers.
We share and discuss such examples.
Finally, our hybrid approach---which selectively employs LLMs when
    conventional methods fail---achieves higher performance than
        individual methods.

\section{Related Work}
\subsection{Program by Example}
When input-output examples are provided, algorithms can find patterns and apply the same transformation to new sets of inputs. 
This method lets users specify their intent through demonstration examples rather than explicit programming, making it a powerful tool for automating repetitive and domain-specific tasks.
After the user provides the input-output examples, the system searches through predefined domain-specific operations to replicate the transformation. Instead of considering every possible function, the core of PBE systems is their ability to restrict the search space to a set of logical operations that are likely to be relevant. The synthesis engine, acting as the head of the network, will search for a program that performs the provided transformation, which involves different computational techniques, such as deductive reasoning, where the system identifies consistent patterns, and inductive learning, where it generalizes rules based on input-output mappings. The goal for such a network is to avoid overfitting and focus on simplicity, the ability to produce a program that can handle a broader range of similar inputs.

Previous PBE-driven tools have demonstrated strong capabilities in enabling users to perform complex data transformations through a few example-based interactions~\cite{ml, string, wrex, ringer, flashextract, Raza_Gulwani_2017, dataXformer}. FlashExtract~\cite{flashextract}, a PBE framework integrated into Excel, enables users to extract structured data from semi-structured text with minimal user input, thereby significantly reducing the time required for data extraction tasks. 

\subsubsection{HoloClean}
HoloClean~\cite{Holo} is a data cleaning system that leverages probabilistic modeling to automatically detect and repair errors in structured data. Its framework consists of three main components: error detection, compilation, and data repairing. During compilation, HoloClean generates a probabilistic model in which random variables represent uncertainty over the values in the input dataset. It uses factor graphs to encode the joint probability distribution over these variables, capturing dependencies derived from integrity constraints, statistical correlations, and external signals. To perform statistical learning and inference, HoloClean builds on DeepDive~\cite{deepdive}, a declarative probabilistic inference framework. Finally, HoloClean repairs detected errors by computing the marginal probabilities of candidate values and selecting the most likely ones.
\subsubsection{Prose}
\begin{figure}[t]
  \centering
  \includegraphics[width=.8\linewidth]{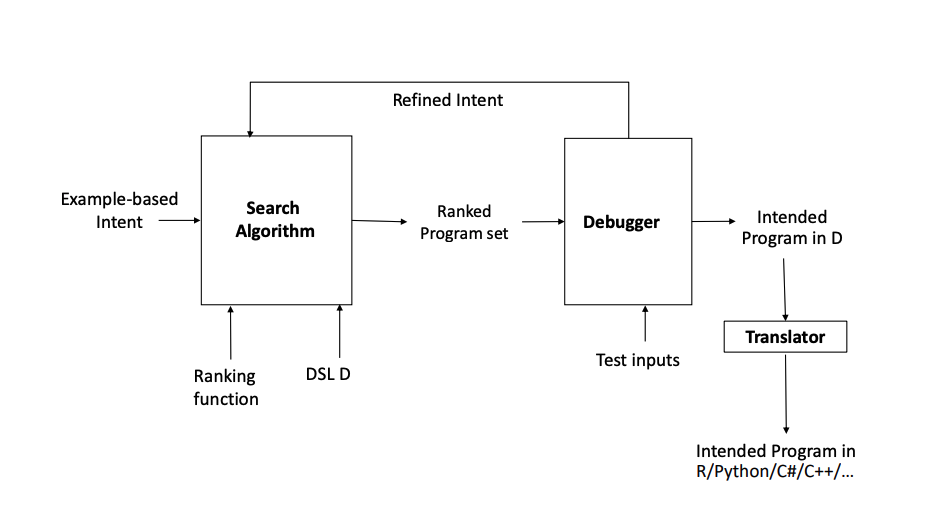}
  \caption{Overview of PBE Architecture\cite{prose}}
  \label{fig:pbe_archi}
\end{figure}

Program Synthesis using Examples (Prose), a research group at Microsoft. The authors described their view of the PBE architecture, which consists of three components: a search algorithm, a ranking strategy, and user interaction models, as shown in Figure \ref{fig:pbe_archi}. The search algorithm is the key to determining if the system is efficient and accurate. A simple search strategy is to go through all the possible
combinations of actions before making the final decision, and maintain a graph structure along the way. This method works well with a small number of operation pools. However, with more complex operations, this bottom-up search approach will have high costs on memory usage and time efficiency.
Thus, they proposed combining Machine Learning (ML) into the program's training process to learn from those mistakes and improve the effectiveness and maintainability of the various PBE components. The combination leads to 8× faster program synthesis, enhances ranking accuracy, and correctly identifies intended programs 50\% more effectively than heuristic-based methods \cite{PLML}. This technique makes PBE more scalable, adaptive, and practical for real-world applications like data wrangling and code transformation. 
\subsubsection{Foofah}
Taking inspiration from the classic A * algorithm \cite{heuristic_algo}, Foofha proposed a heuristic search algorithm to synthesize data transformation. To find a path in a graph, the A* algorithm calculates the cost $f(x) = g(n)+h(n)$, where $g(n)$ is the cost to reach state n from the initial state, and $h(n)$ is the heuristic function calculate the approximate cost of the cheapest path from state n to the goal state, and chose the state with the minimum $f(x)$ to expand. In the case of PBE, instead of focusing on the shortest path, correctness, and readability are more important, the cost is defined by the minimum number of data transformation operations needed from one state to another. With the difference in goal in mind, first, a greedy algorithm to approximate Table Edit Distance(TED) was created: 
\begin{equation}
    TED(T_1, T_2) = \min_{\{p_1, \dots, p_k\} \in P(T_1, T_2)} \sum_{i=1}^{k} \text{cost}(p_i)
\end{equation}
TED calculated the minimum total cost of table edit operations needed to transform from Table 1 to Table 2. Observing frequent simultaneous edits of adjacent cells, TED was further refined into the Table Edit Distance Batch (TED Batch), effectively capturing these grouped operations. With their proposed system Foofah results in an interaction time that is 60\% faster than its predecessors in each test, allowing the users to complete both data syntactic and layout transformation.

Although PBE has demonstrated remarkable capabilities, there are still challenges. These include ensuring it functions properly, handling unclear input patterns, and providing sufficient user interaction. Future work could improve probabilistic models for more reliable program inference. It could also include adding ways for users to provide feedback and creating more effective domain-specific languages (DSLs) to enhance its performance in various situations.

\subsection{Large Language Model}
In recent years, large language models have been extensively studied under the direction of various optimization techniques, aiming to enhance their efficiency and accuracy across different types of tasks. Among these, prompt engineering, fine-tuning, and retrieval-argument generation (RAG) have gained the most popularity and have become the main approaches to enhance the model's ability. Those methods enable models to be more closely tailored to specific tasks, improve accuracy, and reduce computational costs. This section focuses on prompt engineering, covering the key ideas and different approaches employed in this work.
\begin{figure}[t]
  \centering
  \includegraphics[width=.8\linewidth]{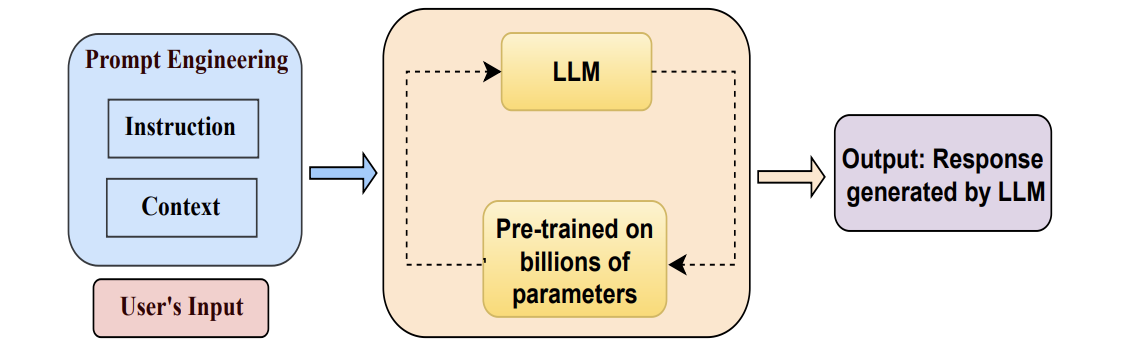}
  \vspace{-2mm}
  \caption{Prompt Engineering Pipeline~\cite{attention}}
  \label{fig:prompt_eng}
\end{figure}

\subsubsection{Zero-shot prompting:} refers to the model's ability to solve the user query without needing examples during inference, eliminating the need for carefully crafted prompts~\cite {radford2019language}. Instead, it depends heavily on the model's pre-training data and its own ability to understand and reason what the user needs. 

\subsubsection{Chain-of-Knowledge (CoK) Prompting:}
LLMs are trained with billions of data points covering a wide range of topics, which may confuse and hallucinate what specific techniques to use. Inspired by human problem-solving, Chain of Knowledge (CoK)~\cite{CoK} systematically breaks down intricate tasks into well-coordinated steps. Engage in a dynamic knowledge adaptation phase, collecting evidence from various sources, such as internal knowledge base, external databases, and the given prompt.


\subsubsection{Multi Turn Reasoning:} 
To reduce the hallucinations that exist in the LLMs responses, multi turn reasoning~\cite{multiturn,rlmultiturn} approach was used. This includes the model first generating an initial solution, then the verification step to check if the original responses sufficiently solve the users' request, and lastly, producing a revised response if necessary. By verifying its own work through this multi-step process, the LLM enhances its logical reasoning ability and reduces hallucination errors. Focused verification step help models identify and correct their inaccuracies.

\section{Experiment Design}

To better understand the capabilities of LLMs, we designed a series of experiments exploring various proposed approaches by systematically combining different dimensions, such as prompt strategies, external knowledge, and hybrid methods. All prompts and datasets used in the experiments are available in our GitHub repository. 

\subsection{Evaluating LLMs on PBE tasks}
We aimed to evaluate the capability of current LLMs in performing PBE tasks related to data transformation. We conducted initial experiments, one shot no knowledge prompt approach, using several baseline models, including Llama, Gemini, Claude, and GPT. The results showed that GPT achieved the highest accuracy at 79.7\%, outperforming Llama-2-13B (17.21\%), Gemini-1.5-pro (47.05\%), and Claude 3.5 (71.41\%). Motivated by GPT’s superior performance, we chose to explore methods to enhance its effectiveness in this domain further.

\subsection{Approach 1: Large-Language Model}
To maintain fairness, all tests were performed using GPT-4o's API call on chat completion under identical settings across all runs.

\subsubsection{Dimension 1: One-shot vs Multiple Tries}
\hfill\\
One-Shot: This vanilla approach prompts the LLM one time with the full task context, including the example input and expected output, and asks it to generate a transformation function that performs the required data manipulation. In this setting, the LLM is expected to synthesize the correct transformation logic in one pass, without any feedback loop or iterative refinement. The goal of this setup is to evaluate the model's ability to generalize and reason correctly with minimal external guidance or opportunity for correction.

Multiple Tries: We also examine the Large Language Model's ability to discover mistakes and whether they are able to learn from its own mistakes and correct itself. To explore this, we proposed the multi-turn verification method. Unlike conventional multi-step approaches, where LLMs see the same question a set number of times, this method provides different prompts at each iteration. And let the LLM utilize the previous responses to identify error based on the output of its prior responses, similar to the chain of verification prompting. 

The process begins with the LLM generating an initial function that it believes will reproduce the data transformation. This function is then executed locally on the example input dataset. If the generated output does not match the example output, two separate experiments are tested: 
\textbf{(a)} The model is provided with its previously generated code, output, and example output. And let the model generate the code again.
\textbf{(b)} An extra error verification step is added to this approach. Where the generated and example outputs are passed to a separate GPT4-o model, specifically prompted to identify high-level structural errors. Then, pass the list of fixes and the previous chat history back to the original model and generate the code again.
Finally, after iterating through this verification loop, the LLM produces the final output, which ideally should successfully reproduce the data transformation on the test dataset.

\subsubsection{Dimension 2: No Knowledge vs Extra Knowledge}
\hfill\\
No Knowledge base prompt: The naive no-knowledge prompt includes the input and output list and a short description of the task the LLM needs to perform. To streamline the process, the test input list was also provided as part of the prompt and asked the LLM to simply put the test list as the function input argument, ensuring a more structured and efficient response.

Knowledge prompt: A list of simple programming functions are provided as part of the prompt. Similar to how Foofah provided the list of operations for its model to search and create an action plan, we used Foofah’s operation set as a starting point—since it covers most commonly seen data transformation techniques—and adapted it to be Python 3 compatible while constraining the search space for efficiency. However, unlike previous conventional PBE programs, which combine multiple sub-programs into a step-by-step series, we asked the LLMs to learn from the provided information and generate a program that performs the transformation in one single function call.
This extra information helps the model reduce the search space and provides a better understanding of the task, and guides it to focus more effectively on data transformation and Python code generation.

\subsection{Approach 2: Search-LLM Hybrid}

LLMs show strong reasoning ability and are great at solving complex, previously unseen tasks. While traditional machine learning models, on the other hand, always bring stability and efficiency. With the advantages of both models in mind, we explored the third method: the hybrid model. In this approach, LLM will take charge when the traditional PBE model struggles to create the perfect program, which might be caused by the example's complexity or by an unseen example dataset in which the PBE model cannot effectively reproduce the transformation. 
By combining these two methodologies, we aim to enhance adaptability and improve performance in challenging data transformation tasks. Moreover, this strategy leverages the strength of both approaches while ensuring that the system maintains robustness when seeing unfamiliar task cases.

\subsection{Experiment Setup}
\paragraph{Baseline:}
Two program synthesis models built for Programming by Example (PBE) in data transformation tasks were used as the baseline comparison. 
\begin{itemize}
    \item Foofah~\cite{foofah}: A PBE system that targets solving both syntactic and layout data transformations. Proposed the program synthesis as a search problem in a state space graph and a heuristic search approach based on their proposed class A* algorithm to synthesize the program.
    \item Prose~\cite{prose}: a research group in Microsoft led by Sumit Gulwani. They were also the first group of people to start looking into the field of Programming by Example.
\end{itemize}
The PBE architecture consists of three components: a search algorithm, a ranking strategy, and user interaction models. The search algorithm is the key to determining if the system is efficient and accurate. A simple search strategy is to go through all the possible combinations of actions before making the final decision, and maintain a graph structure along the way. 

\paragraph{Dataset:}
In this work, two distinct datasets are utilized to evaluate LLMs' capacity to comprehend the underlying data structure and relationships. The first dataset, corrected version shared by Foofah, initially contains the input data in a non-relational format, which, after transformation, will result in a relational output table. 

\begin{itemize}
    \item Foofah~\cite{foofah}: Published its dataset as a combination of the previous datasets, ProgFromEx~\cite{progfromex},
    Wrangler~\cite{wrangler}, Potter’s Wheel (PW)~\cite{pottersw} and Proactive
    Wrangler (Proactive)~\cite{Proactivew} with its contribution, containing 50 test scenarios in total. In the Foofah dataset, each test scenario contains five sub-test files, numbered 1 through 5, indicating the number of data records selected from the raw data as the example input and output that are passed into the program.
    \item Prose~\cite{prose}: used has a semi-structured format. Unlike the relational dataset, semi-structured data does not adhere to a strict schema, presenting a distinct set of challenges for LLMs. The Prose dataset includes tasks involving string transformations and Excel-like data manipulation, often formatted in JSON or XML, making it a useful benchmark for evaluating model performance on real-world, flexible data representations.
    
\end{itemize}

\paragraph{Metrics:}
Exact match accuracy was used for evaluation, comparing the transformed table produced by the model’s generated function directly with each test case’s ground truth output table. For each test case, we compared the resulting table row by row to check whether each data point matched precisely with the corresponding value in the expected output.
Then, the accuracy of each row was calculated based on the percentage of data points in that row that were correct; the overall accuracy for each test dataset is the average accuracy of all rows. Unlike binary exact match, this approach offers a more detailed assessment by capturing partial accuracy inside every sample. Hence, it presents a fuller picture of the model's ability. To report the final accuracy we used the weighted average, as each dataset contains a different distribution of transformation types, and using a simple macro-average could overemphasize datasets with rare or easier transformations. By applying a weighted average, we ensure that the contribution of each dataset reflects its size and diversity, providing a fair and balanced overall evaluation.

\section{Evaluation Results}

\begin{table*}[ht]
\caption{
Accuracy on Foofah and Prose Dataset
}
\vspace{-2mm}
\centering
\small
\begin{tabular}{l r r r r r r }
\toprule
\multirow{2}{*}{\textbf{Approach}} & \multicolumn{5}{c}{\textbf{Dataset}} &
    \multirow{2}{*}{\textbf{Overall (Weighted Avg.)}}
    \\  \cline{2-6}\noalign{\vskip 2pt}
& ProgEx~\cite{progfromex} & Wrgler~\cite{wrangler} & Potter~\cite{pottersw} & Proact~\cite{Proactivew} & Prose~\cite{prose} & \\
\midrule
Prose           &  0.139  &  0.286  &  0.183 & 0.156  & \textbf{0.949} & 0.473         \\
Foofah           & 0.689   & 0.667 & \underline{0.891}   & \textbf{1}   & 0.300  & 0.571        \\
Gemini-3.5       & 0.790   & 0.670        & 0.793     & \textbf{1}    &   0.566    & 0.714    \\
Llama2-13B      & 0.068   & 0.433        & 0.135     & 0.043    &   0.257    & 0.172    \\

Gemini-1.5-pro       & 0.068   & 0.417        & 0.677     & 0.400    &   0.530    & 0.471    \\
GPT-4o + One-Shot       & 0.368   & 0.733        & 0.886     & 0.770    &   0.755    & 0.797    \\
GPT-4o + Multi-tries (a) &  \textbf{0.813} &  0.827 & 0.874  & 0.876  &  0.858   & 0.786 \\
GPT-4o + Multi-tries (b) & 0.787  & \textbf{0.870}  & 0.857  & \underline{0.943}  &   0.873  & \underline{0.846}   \\
GPT-4o + One-Shot +  Extra kg       & 0.759               & 0.713        & 0.833     & 0.880     &   0.731   & 0.766   \\
GPT-4o + Multi-tries (a)+  Extra kg &  0.778 & 0.763  &  0.814 & 0.890  &  0.860  & 0.827   \\
GPT-4o + Multi-tries (b)+  Extra kg & 0.718  &  \underline{0.867} &  0.794 & 0.933  &  0.878 &0.811  \\
Foofah + GPT-4o + Multi-tries &  \underline{0.802} & \textbf{0.870}  &  \textbf{0.904} & \textbf{1}  &  \underline{0.883} & \textbf{0.863} \\
\bottomrule
\end{tabular}
\label{tbl:accy}
\end{table*}

Since running the same task multiple times is often unrealistic in real-world applications, where users expect reliable results in a single attempt. Therefore, instead of running the model multiple times and averaging the accuracy for each test case,  we evaluated the correctness of the generated programs from each approach based on a single execution per model across the two datasets. 
\begin{itemize}
    \item How would different dimensions from approach 1 influence the accuracy
    \item How well these models generalize across domains and datasets
\end{itemize}

\subsection{LLMs Benefit From Knowledge and Tries}

Baseline: To establish a baseline for comparison, we consider the accuracy of Foofah and Prose, which are traditional search-based program synthesis methods. Foofah achieves a weighted average accuracy of 0.571, while Prose scores 0.473. These scores serve as benchmarks for evaluating the effectiveness of newer LLM-based approaches.

One-shot vs Multiple Tries:  Comparing GPT-4o + One-Shot (0.797) to GPT-4o + Multi-tries (0.786), we observe that one-shot performance is slightly better in this instance. However, the difference is minimal, and multi-try methods tend to be more consistent across datasets. When combined with extra knowledge, the multi-try setup (best at 0.827) clearly outperforms its one-shot counterpart with extra knowledge (0.766), demonstrating the value of multiple attempts in more informed settings. With the two variants from Multi-tries, they both deliver competitive performance. Specifically, variant (b) achieves slightly higher peak scores (0.943) compared to (a) (0.876), indicating that variant (b) is superior at achieving high-quality results in certain instances. However, variant (a) demonstrates more stable consistency across other metrics, suggesting it provides balanced outcomes.

No Knowledge vs Extra Knowledge: To evaluate the effect of external knowledge, we compare setups without and with extra knowledge. For GPT-4o + Multi-tries, adding extra knowledge improves performance from 0.786 to 0.827. Similarly, GPT-4o + Multi-tries with extra knowledge achieves consistently high accuracy across datasets, indicating that domain or dataset-specific guidance significantly enhances synthesis outcomes. Notably, combining Foofah with GPT-4o + Multi-tries further boosts the weighted average accuracy to 0.863, the highest among all tested approaches.

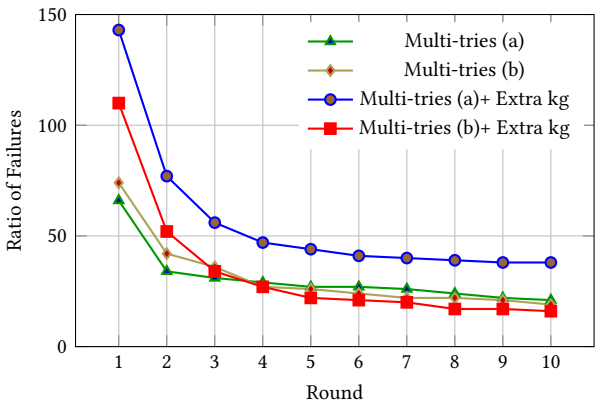
\begin{figure}[t]
    \centering
\begin{tikzpicture}
\begin{axis}[
    width=\linewidth,
    height=6cm,
    xlabel={Round},
    ylabel={Ratio of Failures},
    xtick={1,2,3,4,5,6,7,8,9,10},
    ymin=0, ymax=150,
    grid=both,
    legend style={
        draw=none,
        fill=white,
        font=\small,
        at={(0.98,0.98)},
        anchor=north east
    },
    label style={font=\small},
    tick label style={font=\small},
]

\addplot+[mark=triangle*, color=green!60!black, thick] 
    coordinates {(1,66) (2,34) (3,31) (4,29) (5, 27) (6,27) (7, 26) (8, 24) (9, 22) (10,21)};
\addlegendentry{Multi-tries (a)}

\addplot+[mark=diamond*, color=yellow!60!black, thick] 
    coordinates {(1,74) (2,42) (3,36) (4,27) (5, 26) (6,24) (7, 22) (8, 22) (9, 21) (10,19)};
\addlegendentry{Multi-tries (b)}

\addplot+[mark=*, color=blue, thick] 
    coordinates {(1,143) (2,77) (3,56) (4,47) (5, 44) (6,41) (7,40) (8,39) (9,38) (10,38) };
\addlegendentry{Multi-tries (a)+ Extra kg}

\addplot+[mark=square*, color=red, thick] 
    coordinates {(1,110) (2,52) (3,34) (4,27) (5, 22) (6,21) (7, 20) (8, 17) (9, 17) (10,16)};
\addlegendentry{Multi-tries (b)+ Extra kg}

\end{axis}
\end{tikzpicture}
\vspace{-2mm}
\caption{Trends in code generation failures across iterative rounds of the loop process. Failures are defined as either a mismatch with the intended data transformation logic or execution errors. The plot demonstrates a consistent decrease in failure rate with increased iteration.}
\end{figure}

Hybrid Method: Combining traditional systems like Foofah with GPT-4o Loop Verification (a) Base Prompt resulted in 0.863 accuracy overall. It performed well on Proact (1.0) and Prose (0.883), but still faced challenges with intricate datasets.

\subsection{LLM Could Complement Conventional PBE}

One limitation of the traditional PBE system is that it is all built within the knowledge of a specific domain. Even the overall task may fall under the general area of data transformation; however, the model's ability is limited by the program provided to the system and the rules available to it. As a result, the previous PBE systems often struggle when seeing tasks outside their designated area, leading to limited usability and a lack of flexibility when adapting to new or unfamiliar transformation patterns. 

However, this is not a problem for large language models. With billions of data points being passed to different models for the training process, the model is designed to solve a wide range of tasks with its extensive world knowledge. This nature enables LLMs to handle unseen or complex data transformation tasks with fewer human-enforced guidelines, as the table below indicates. For instance, one test case from the Prose dataset~\cite{prose} involves mapping language names to their corresponding ISO 639-1 codes, such as ["Arabic = ar"], ["Basque = eu"]. GPT-4o is able to solve this correctly, whereas Foofah, constrained by its limited domain knowledge, fails to generalize to such examples.

\begin{table}[h]
    \centering
    \caption{Evaluation result on cross-domain datasets}
    \vspace{-2mm}
    \begin{tabular}{|c|c|c|}
        \hline
          & Foofah Dataset & Prose Dataset \\ \hline
        \% test cases passed to LLM & 15.88\% & 78.95\%  \\ \hline
        \% test cases solved by LLM & 48.65\% & 45.19\% \\ \hline
    \end{tabular}
    \label{tab:cross-domain results}
\end{table}
These differences become particularly evident when comparing the performance of LLM vs. PBE systems, Foofah and Prose, across various datasets. As shown in the table \ref{tbl:accy}, Prose did remarkably well on its own dataset, achieving around 90\% accuracy. However, it experienced a considerable performance drop on Foofha's dataset, highlighting its limitations in generalizing across domains, with unseen transformation types or dataset structures. Similarly, Foofah performs well on its domain dataset but decreases its performance when evaluated on the Prose dataset.

In contrast, LLM, with only the base prompt and no other human intervention or task-specific tuning, performs relatively well on both datasets, achieving 82.42\% on the Foofah dataset and 75.53\% on the PROSE dataset on transformation text. With the highest accuracy achieved using the hybrid method, LLM has successfully addressed the domain-specific limitations of the traditional PBE system.

\section{Discussion}

\begin{figure}[t]
  \label{tab:cases-ambiguity}
  \centering
  \setlength{\tabcolsep}{2pt}
  \footnotesize

  \begin{subfigure}[t]{0.9\linewidth}
    \centering
    \begin{tabular}{cc}
      \begin{tabular}{|c|c|c|c|c|c|}
        \hline
        \multicolumn{6}{|c|}{\textbf{Example Input}} \\ \hline
        001-001 & 1 &  &  &  & \$-         \\ \hline
        001-001 & 2 &  &  &  & \$-         \\ \hline
        001-001 & 3 &  &  &  & \$7,664.25  \\ \hline
        001-001 & 4 &  &  &  & \$-         \\ \hline
      \end{tabular}
      \qquad
      \begin{tabular}{|c|c|c|c|c|}
        \hline
        \multicolumn{5}{|c|}{\textbf{Example Output}} \\ \hline
        001-001 & \$- & \$- & \$7,664.25 & \$- \\ \hline
      \end{tabular}
    \end{tabular}
    \caption{Example transformation in prompt}
  \end{subfigure}

  \vspace{2mm}

  \begin{subfigure}[t]{0.9\linewidth}
    \centering
    \begin{tabular}{cc}
      \begin{tabular}{|c|c|c|c|c|c|}
        \hline
        \multicolumn{6}{|c|}{\textbf{Test Input}} \\ \hline
        001-001 & 2 &  &  &  & \$-         \\ \hline
        001-001 & 4 &  &  &  & \$-         \\ \hline
        001-001 & 6 &  &  &  & \$-         \\ \hline
        001-001 & 8 &  &  &  & \$-         \\ \hline
        001-001 & 9 &  &  &  & \$7,664.25  \\ \hline
      \end{tabular}
      \qquad
      \begin{tabular}{|c|c|c|c|c|c|}
        \hline
        \multicolumn{6}{|c|}{\textbf{Ground Truth}} \\ \hline
        001-001 & \$- & \$- & \$- & \$- & \$7,664.25 \\ \hline
      \end{tabular}
    \end{tabular}
    \caption{Expected, Ground-truth transformation}
  \end{subfigure}

  \vspace{2mm}

  \begin{subfigure}[t]{0.9\linewidth}
    \centering
    \begin{tabular}{cc}
      \begin{tabular}{|c|c|c|c|c|c|}
        \hline
        \multicolumn{6}{|c|}{\textbf{Test Input}} \\ \hline
        001-001 & 2 &  &  &  & \$-         \\ \hline
        001-001 & 4 &  &  &  & \$-         \\ \hline
        001-001 & 6 &  &  &  & \$-         \\ \hline
        001-001 & 8 &  &  &  & \$-         \\ \hline
        001-001 & 9 &  &  &  & \$7,664.25  \\ \hline
      \end{tabular}
      \qquad
      {\setlength{\tabcolsep}{0.4em}
      \begin{tabular}{|c c c c c c|}
        \hline
       \multicolumn{6}{|c|}{ \textbf{GPT Output}}\\ \hline
         &  &  & \hspace{-1em}\textcolor{red}{no output: indexing error} & & \\ \hline
      \end{tabular}}
    \end{tabular}
    \caption{Actual transformation by generated code}
  \end{subfigure}

  \caption{The model incorrectly generalizes from the example by treating the second column as a sequential index, leading to a list assignment index error when test data lacks a continuous sequence.}
  \label{fig:amb_input}
\end{figure}

In this section, we analyze two of the most common and impactful failure cases encountered when using LLM-generated code for data transformation tasks. Despite their overall effectiveness, LLMs exhibit systematic weaknesses under certain conditions. The first failure case, driven by ambiguity in the input data, caused 87\% of the test cases to fail in this area, highlighting the model's tendency to overfit to patterns in the examples rather than generalizing the underlying logic. The second case, which requires multi-dimensional reasoning and contextual understanding, resulted in a 60\% failure rate, showing that LLMs often struggle to infer implicit constraints or real-world expectations.

\subsection{LLM failed due to ambiguity.}

We highlight a failure case where the LLM-generated code does not generalize correctly due to ambiguity in the input data. Specifically, the model overfits to patterns present in the example dataset, making incorrect assumptions about the structure of the input. This illustrates a common challenge in LLM-based code generation: when the intent is underspecified or when misleading patterns exist in the examples, the model may infer incorrect logic that does not hold in broader contexts.

As shown in the example in figure \ref{fig:amb_input}, the model should realize that the transformation only extracts the last entry from each row and combines it into a single row with the same 'item-id' (the first element in the row). Instead, because the second element in each row in the example starts with one and has an increment of 1 in the following rows, the model mistakenly interprets it as an index and applies this logic to the program. While this works correctly in the example datasets, the program fails the test cases. In the test case, the second item in each row is no longer a continuous sequence, which causes the list assignment index out-of-range error.

\subsection{LLM failed due to Multi-Dimensional Tasks}

Some transformation tasks go beyond simple structural manipulation and require reasoning over multiple domains, such as temporal context, commonsense knowledge, or domain-specific conventions. These tasks pose a greater challenge for LLMs, which may correctly parse and transform text but fail to infer implicit meanings or constraints. For example, as shown in the figure \ref{fig: author} below, the task is to extract the year of death from the list of authors. 

When examining the data, it is an instinctive assumption for humans that a value like "65" in the table refers to the year 1965 or at least any century before the current year. However, when looking at the program result from the LLM-generated code, GPT-4o was able to slice the string successfully and extract the year from the original input data. Still, it was unable to reason that the data refers to the author's passing away, and the date should not be in the current year. These examples and others from the PROSE dataset illustrate a limitation of the GPT-4o model in that it can handle structure-wise transformations well. However, they may struggle with tasks that require them to think of another step and presume the correct meaning of the text.

\begin{figure}[t]
  \centering
  
  \setlength{\tabcolsep}{2pt}
  \footnotesize

  \begin{subfigure}[t]{0.5\textwidth}

    \begin{tabular}{cc}

      \begin{tabular}{|c|c|c|c|c|}
        \hline
        \multicolumn{5}{|c|}{\textbf{Example Input Table}} \\ \hline
        21-May-00 & 1973 & Living & 6-Nov-62 & 5 December 1870 \\ \hline
      \end{tabular}
      &
      \begin{tabular}{|c|c|c|c|c|}
        \hline
        \multicolumn{5}{|c|}{\textbf{Example Output Table}} \\ \hline
        2000 & 1973 & 2025 & 1962 & 1870 \\ \hline
      \end{tabular}
    \end{tabular}
    \caption{Example transformation in prompt}
  \end{subfigure}
  \hfill

  \begin{subfigure}[t]{0.5\textwidth}
    \centering
    \begin{tabular}{cc}
      \begin{tabular}{|c|c|c|c|c|}
        \hline
        \multicolumn{5}{|c|}{\textbf{Test Input Table}} \\ \hline
        Living & 2-Dec-65 & 1 December 1848 & 1984 & 28-Nov-68 \\ \hline
      \end{tabular}
      &
      \begin{tabular}{|c|c|c|c|c|}
        \hline
        \multicolumn{5}{|c|}{\textbf{Ground Truth Table}} \\ \hline
        2025 & 1965 & 1848 & 1984 & 1968 \\ \hline
      \end{tabular}
    \end{tabular}
    \caption{Expected, Ground-truth transformation}
  \end{subfigure}
  \hfill

  \begin{subfigure}[t]{0.5\textwidth}
    \centering
    \begin{tabular}{cc}
      \begin{tabular}{|c|c|c|c|c|}
        \hline
        \multicolumn{5}{|c|}{\textbf{Test Input Table}} \\ \hline
        Living & 2-Dec-65 & 1 December 1848 & 1984 & 28-Nov-68 \\ \hline
      \end{tabular}
      &
      \begin{tabular}{|c|c|c|c|c|}
        \hline
        \multicolumn{5}{|c|}{\textbf{GPT Output Table}} \\ \hline
        2025 & \textcolor{red}{2065} & 1848 & 1984 & \textcolor{red}{2068} \\ \hline
      \end{tabular}
    \end{tabular}
    \caption{Actual transformation by generated code}
  \end{subfigure}
  \caption{GPT-4o extracts dates accurately but lacks commonsense reasoning, misinterpreting past events as future ones.}
  \label{fig: author}
\end{figure}

\section{Conclusion}
In this work, we examine the logical reasoning and code generation ability of LLMs through PBE tasks focused on tabular formated data transformation. Three approaches were proposed that do not alter the model’s structure and minimize user effort to enhance the model’s performance on this specific task. As the experiment result indicates, all of them were able to outperform Foofah’s PBE system performance with the highest accuracy, reaching 86.3\%. These approaches demonstrate that with carefully designed prompting strategies, LLMs can perform structured tasks more effectively without any fine-tuning or architecture modification.
As the target users for such methods are end users who may not have technical backgrounds, the methods we mentioned are a great fit as they do not require modifying the model structure or setting up additional tools.
They also highlight how adaptable modern LLMs can be with different type of tasks when paired with the proper reasoning framework and problem decomposition strategy. There still remain promising directions for future research, as discussed earlier, to further enhance LLM performance with tabular data on more complex tasks such as PBE, including handling noisier data, multi-step reasoning, and more generalized transformation goals.

\bibliographystyle{ACM-Reference-Format}

\end{document}